\documentclass{emulateapj}
\usepackage{apjfonts}
\usepackage{url}
\usepackage{hyperref}
\usepackage{amsmath}

\newcommand{\sat}[1]{\it\uppercase{#1}\rm}
\newcommand{\fig}[1]{Fig.~\ref{#1}}

\newcommand{\speed}[1]{#1 km~s${}^{-1}$}
\newcommand{\aspeed}[1]{$\sim\,$#1 km~s${}^{-1}$}
\newcommand{\accel}[1]{#1 m~s${}^{-2}$}

\newcommand{\rsun}[1]{${#1}\,R_\odot$}

\begin{document}

\shorttitle{Early Evolution of CME} %

\shortauthors{Liu}

\title{Early Evolution of An Energetic Coronal Mass Ejection And Its Relation to EUV Waves} 

\author{Rui Liu, Yuming Wang, \& Chenglong Shen}

\affil{CAS Key Laboratory of Geospace Environment, Department of Geophysics and Planetary Sciences, University of Science and
Technology of China, Hefei 230026, China}
\email{rliu@ustc.edu.cn}

\begin{abstract}
We study a coronal mass ejection (CME) associated with an X-class flare, whose initiation is clearly observed in low corona with high-cadence, high-resolution EUV images, providing us a rare opportunity to witness the early evolution of an energetic CME in detail. The eruption starts with a slow expansion of cool overlying loops ($\sim\,$1 MK) following a jet-like event in the periphery of the active region. Underneath the expanding loop system a reverse S-shaped dimming is seen immediately above the brightening active region in hot EUV passbands. The dimming is associated with a rising diffuse arch ($\sim\,$6 MK), which we interpret as a preexistent, high-lying flux rope. This is followed by the arising of a double hot channel ($\sim\,$10 MK) from the core of the active region. The higher structures rise earlier and faster than lower ones, with the leading front undergoing extremely rapid acceleration up to 35 km~s$^{-2}$. This suggests that the torus instability is the major eruption mechanism and that it is the high-lying flux rope rather than the hot channels that drives the eruption. The compression of coronal plasmas skirting and overlying the expanding loop system, whose aspect ratio $h/r$ increases with time as a result of the rapid upward acceleration, plays a significant role in driving an outward-propagating global EUV wave and a sunward-propagating local EUV wave, respectively.

\end{abstract}


\section{Introduction}
CMEs are the most important space weather-relevant events on the Sun, in close association with solar flares and prominence eruptions. The three eruptive phenomena are hence suggested to be different manifestations of a single physical process, which involves the large-scale disruption and restructuring of the coronal magnetic field \citep{forbes00,lin03}. 
 
Most often CMEs are studied at high altitudes using coronagraphs \citep{wh12}, although their formation and initiation take place in low corona. The reason is twofold: on one hand, the CME progenitor is often hardly discernible in the traditional EUV passbands (e.g., 171 and 195~{\AA}) without the presence of a dense prominence; on the other, the observational coverage of low corona, in terms of cadence, resolution, and temperature discrimination ability, has not been sufficient to capture in detail the early impulsive acceleration of CMEs, until the introduction of the Solar Dynamic Observatory \citep[SDO;][]{pesnell12}. Recent SDO observations demonstrate that the initiation of a CME is often preceded by the formation a flux rope in low corona, which appears as a twisted hot channel in the Atmospheric Imaging Assembly \citep[AIA;][]{lemen12} 94 and 131~{\AA} passbands \citep[e.g.,][]{liu10,cheng11,zhang12}. \citet{cheng13} concluded from a study of two CMEs that the hot channel serves as a continuous driver in the early acceleration phase since it rises earlier, and at a faster speed, than the CME leading front in low corona. The leading front, transformed from pre-existent coronal loops, often appears as a bubble in cool EUV passbands \citep[e.g.,][]{pvk10,pvs10,cheng13}. 

In this paper, we report on a CME associated with an X-class flare, which has a double hot channel forming in the core of the CME during the early phase. In the sections that follows, we study in detail the initiation and the early evolution of the CME in low corona and its relation to coronal waves and oscillations (Section 2); we then highlight the unique features that are uncovered by the high-cadence, high-resolution data (Section 3). The CME's subsequent evolution in high corona and propagation in interplanetary space as well as its impact on space weather will be covered in a separate study.

\section{Observation \& Analysis}

\subsection{Instruments \& Data Reduction}
The \sat{goes}-class X-3.2 flare takes place on 2013 May 15 at N10E68 in the NOAA active region 11748. It starts at 01:25 UT and peaks at 01:48 UT. The initiation of the CME is clearly observed in EUV by \sat{sdo}/AIA. The six EUV passbands of AIA, i.e., 131~{\AA} (peak response temperature $\log T=7.0$), 94~{\AA} ($\log T=6.8$), 335~{\AA} ($\log T=6.4$), 211~{\AA} ($\log T=6.3$), 193~{\AA} ($\log T=6.1$) and 171~{\AA} ($\log T=5.8$), can be used to calculate the differential emission measure (DEM) in the logarithmic temperature range $[5.5,7.5]$. We utilize the code developed by \citet{plowman13}, which is a fast iterative method based on regularization, using the AIA temperature response functions as basis functions. We use the most recent available AIA temperature response functions (V4 calibration), applying both the CHIANTI fix to account for missing emission lines in the CHIANTI v7.0 database in channels 94 and 131~{\AA} and EVE normalization to give good agreement with full-disk EVE spectral irradiance data. As inputs to the DEM code, the AIA level-1 data are processed to level 1.6 by deconvolving the point spread function using \texttt{AIA\_DECONVOLVE\_RICHARDSONLUCY} before being co-registered to a common center and binned to the same pixel size using \texttt{AIA\_PREP}. The accuracy of the co-alignment is on the order of sub-pixels as confirmed by \texttt{AIA\_COALIGN\_TEST}. 

\begin{figure}
  \centering
  \includegraphics[width=\hsize]{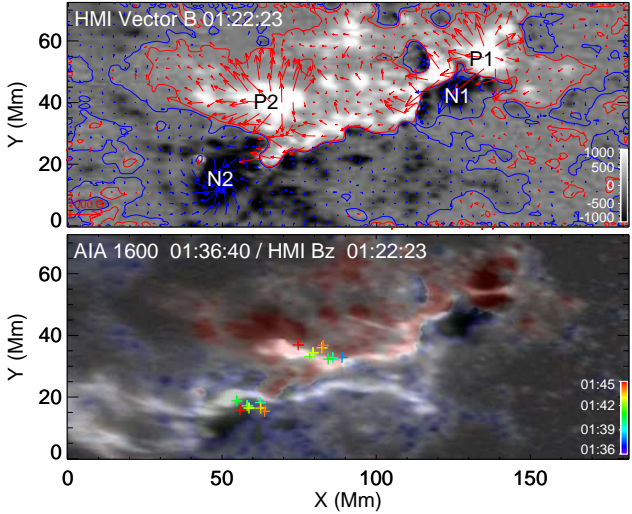}
  \caption{\small Pre-flare field configuration. Top: photospheric $B_z$ distribution, saturated at $\pm1000$ G, is superimposed with contours at $B_z=\pm50$ G, with red (blue) colors indicating positive (negative) polarities. Red (blue) arrows denote the horizontal component of the photospheric field originating from positive (negative) $B_z$. Bottom: an AIA 1600~{\AA} image (black and white) taken at the onset of the nonthermal phase of the flare is blended with the pre-flare $B_z$ map (red and blue); both are remapped with the CEA projection. The centroid positions of HXR footpoint sources at 50--100 keV are marked by plus signs, whose temporal evolution is color coded. \label{hmi}}
\end{figure}

Magnetograms obtained by the Helioseismic and Magnetic Imager \citep[HMI;][]{schou12a,schou12b} onboard \sat{sdo} provide the magnetic context of the source region. A pre-flare vector magnetogram at 01:22 UT is shown in the top panel of Fig.~\ref{hmi}. One can see that the active region has a typical quadrupolar configuration. The four flux concentrations, labeled P1--N1 and P2--N2 in Fig.~\ref{hmi}(a), correspond to two pairs of sunspots. The polarity inversion line (PIL) can be identified where red and blue contours of $B_z=\pm 50$ G contact each other. Note that the magnetogram is remapped with a cylindrical equal area projection (CEA). In the bottom panel, a remapped AIA 1600~{\AA} image taken at the onset of nonthermal hard X-rays (HXRs) is blended with the $B_z$ map, showing that the pair of conjugate flare ribbons are parallel to the major PIL (bottom panel of \fig{hmi}). Superimposed are the centroid positions of the flare footpoints at 50--100 keV (plus signs) given by the Reuven Ramaty High-Energy Solar Spectroscopic Imager \citep[\it RHESSI\rm;][]{lin02}. One can see that the flare footpoints are mainly associated with P2--N2, which gives clues on where magnetic free energy is accumulated and released. 

\subsection{CME Dynamics in Low Corona}

\begin{figure*}
  \centering
  \includegraphics[width=0.95\hsize]{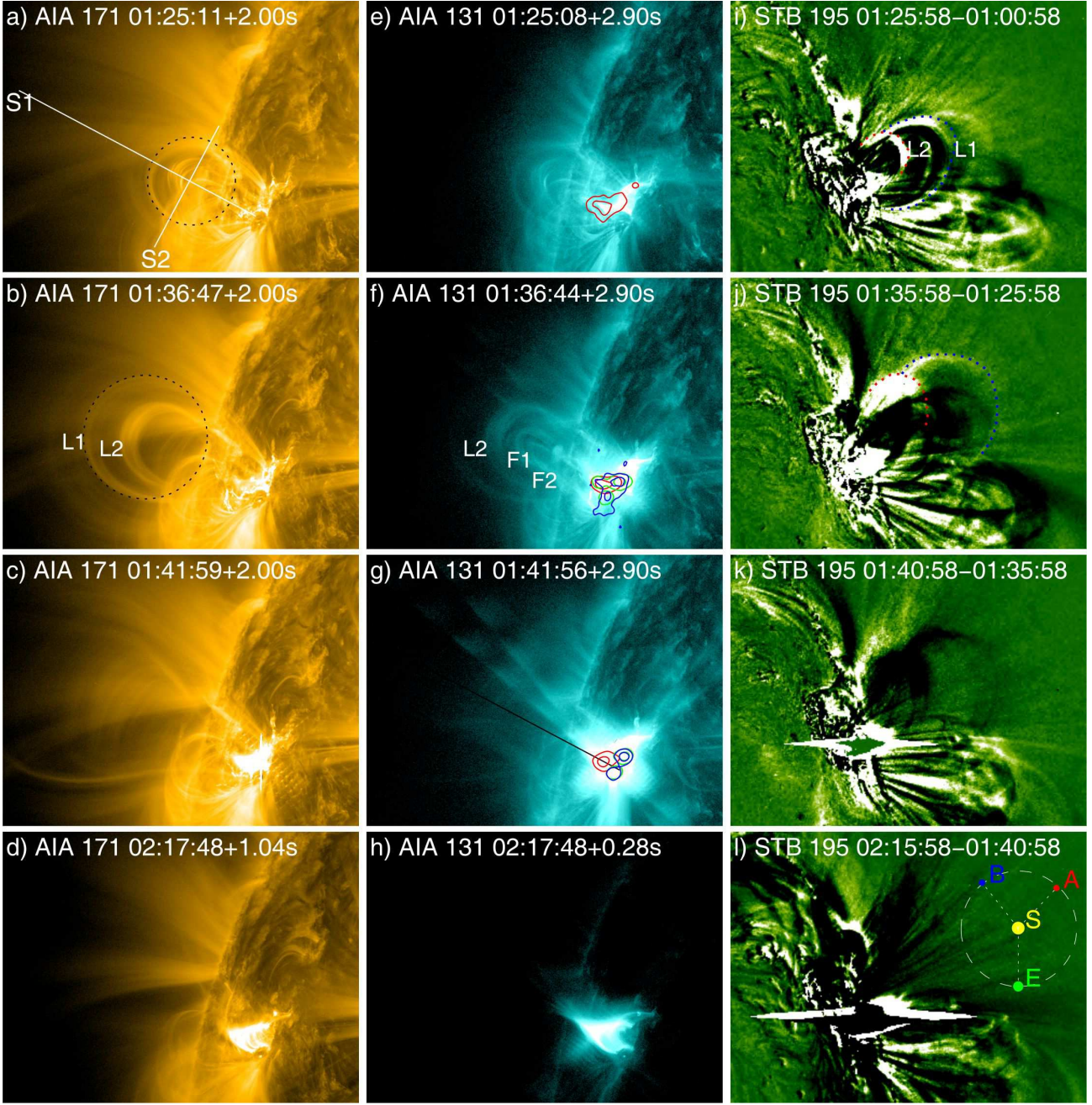}
  \caption{\small Snapshots of the CME observed in low corona in AIA 171 (left) and 131~{\AA} (center) and EUVI-B 195~{\AA} (right). Superimposed contours denote hard X-ray emission at 12--25 keV (red), 25--50 keV (green) and 50--100 keV (blue), with contour levels at 30\% and 70\% peak emission of each individual RHESSI image. Two virtual slits S1 and S2 are indicated in (a). S1 is replotted in (g). L1's geometrical parameters are obtained by fitting visually picked points along L1 with circles, as demonstrated in (a) and (b). Each EUVI image is subtracted by an image acquired earlier. In Panels (i) and (j), blue (red) dotted lines delineate L1 (L2). The inset of Panel (l) indicates the positions of the two STEREO satellites (`A' and `B') with respect to the Sun (`S') and Earth (`E') in the Earth ecliptic plane. An animation of 131~{\AA} images is available online. \label{overview}}
\end{figure*}

\begin{figure*}
  \centering
  \includegraphics[width=\hsize]{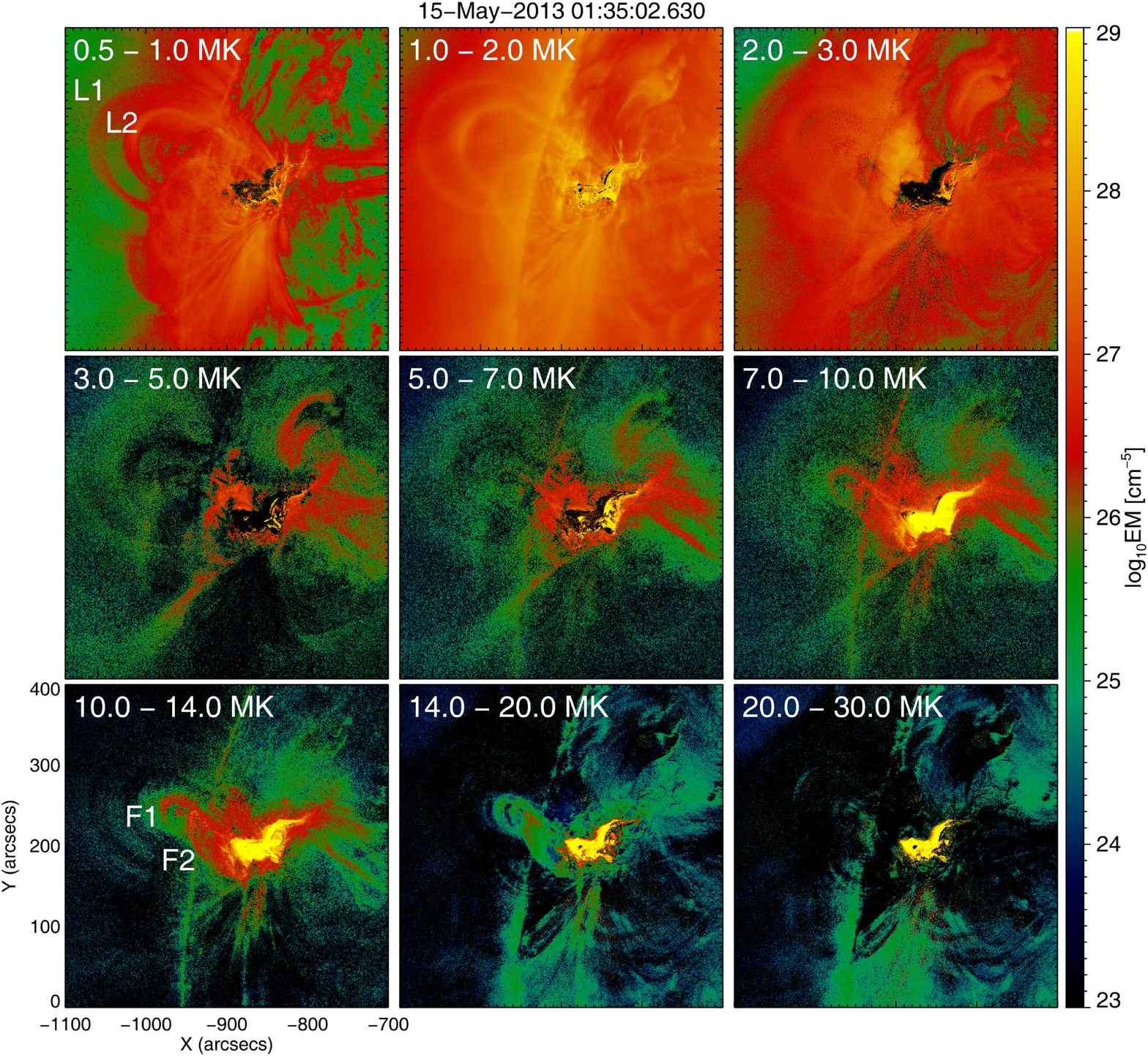}
  \caption{Emission measure in different temperature ranges for the 2013-May-15 flare at 01:35 UT. \label{demdt}}
\end{figure*}

\fig{overview} shows a sequence of coronal images taken in the \sat{SDO}/AIA 171 and 131~{\AA} passbands. The active region is also observed by the Extreme Ultraviolet Imager \citep[EUVI;][]{wuelser04}) onboard the `Behind' satellite (STB hereafter) of the Solar TErrestrial RElations Observatory \citep[STEREO;][]{kaiser08}. Starting from about 01:00 UT onward, a loop system is observed to gradually expand. In 171~{\AA}, two bundles of loops that are clearly seen are labeled `L1' and `L2', respectively, with L2 apparently nested underneath L1 (\fig{overview}(b)), but from \sat{STB}'s perspective (separated from Earth by 141.5 deg in the ecliptic plane; see the inset of Fig.~\ref{overview}(l)), they cross each other, with L2 in front of L1 (\fig{overview}(i) and (j)), suggesting that they are moving outward in different directions. Note that we select L1 and L2 for further analysis because of their superior clarity and coherence to other loops within this expanding loop system. From about 01:20 UT onward, two tangled ``hot channels'' are observed to rise only in 131~{\AA} and 94~{\AA}, labeled `F1' and `F2' in \fig{overview}(f) (see also the animation accompanying \fig{overview}). The DEM analysis (\fig{demdt}) confirms that L1 and L2 are cool (0.5--2 MK) while F1 and F2 are hot ($\sim\,$10 MK). With the rising of these hot channels, the expansion rate of L1 and L2 accelerates, and the lower section of their legs apparently approach each other as a result of the rapid expansion (\fig{overview}(c)). Eventually, the eruption leaves behind a typical post-flare arcade in 171~{\AA} (\fig{overview}(d)) connecting the pair of UV flare ribbons in 1600~{\AA}. Cusp-like flare loops are observed in 131~{\AA} (\fig{overview}(h)) overlying the round-shaped arcade. 

\begin{figure*}
  \centering
  \includegraphics[width=0.85\hsize]{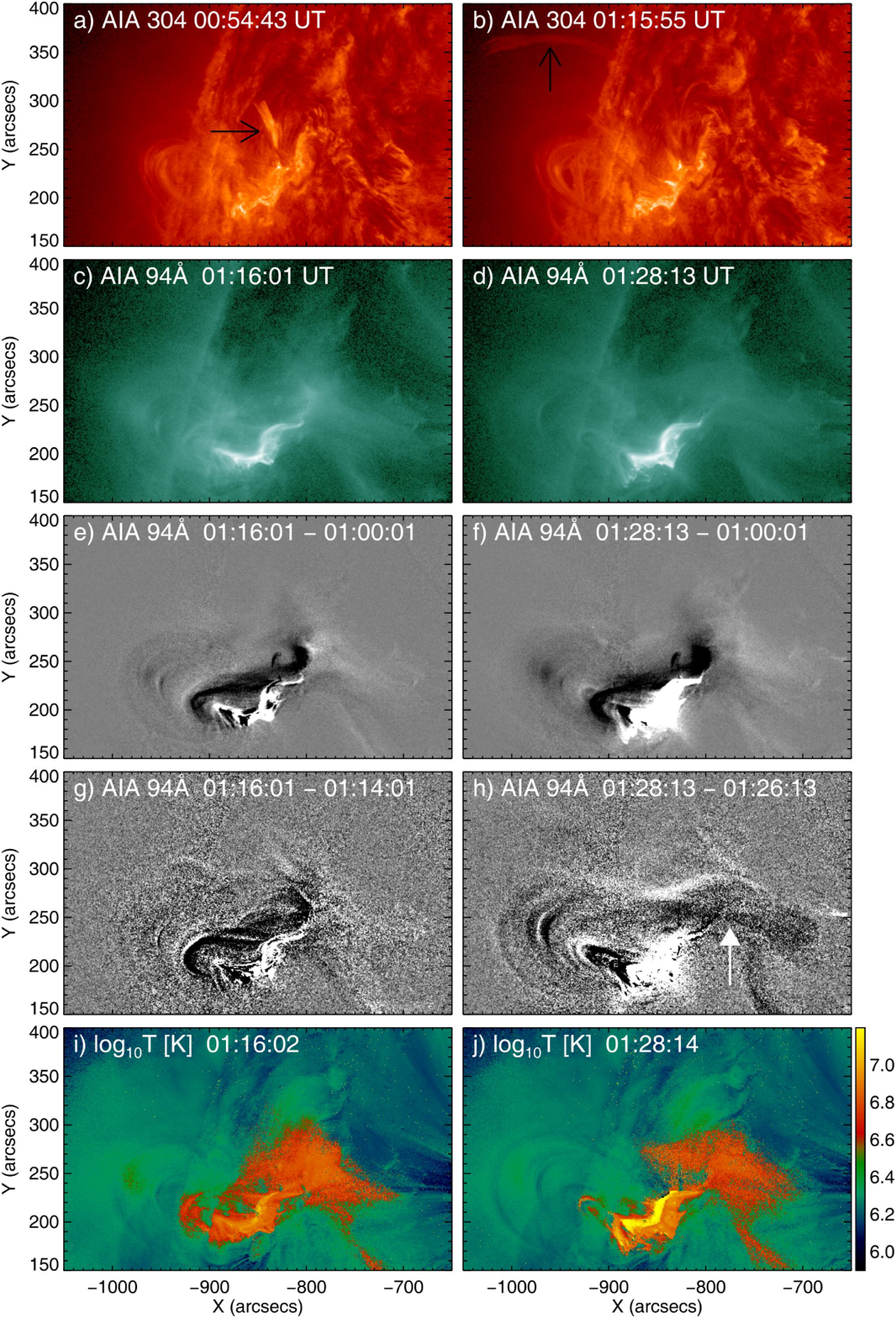}
  \caption{Pre-eruption dynamics. Top panels show a jet-like event in AIA 304~{\AA}. From the second row to bottom displayed are the AIA 94~{\AA} images at 01:16 (left) and 01:28 UT (right), the corresponding base- and running-difference images, and the maps of emission-measure-weighted temperatures obtained from the DEM analysis. \label{init}}
\end{figure*}

Even before the emergence of the two hot channels, one can see a reverse S-shaped dimming immediately above the brightening active region in AIA 94~{\AA} base difference images (Fig.~\ref{init}(e--f)). This dimming is associated with the slow rise of a diffuse, arch-like structure (marked by a white arrow in Fig.~\ref{init}(h)) of about 6 MK (Fig.~\ref{init}(g--h)) overlying the active region, which moves at a similar pace as L1 and L2. A jet-like event starting as early as 00:54 UT is detected in the periphery of the active region in AIA 304~{\AA} (Fig.~\ref{init}(a--b); marked by black arrows), preceding the rising of the arch-like structure.

\begin{figure*}
  \centering
  \includegraphics[width=\hsize]{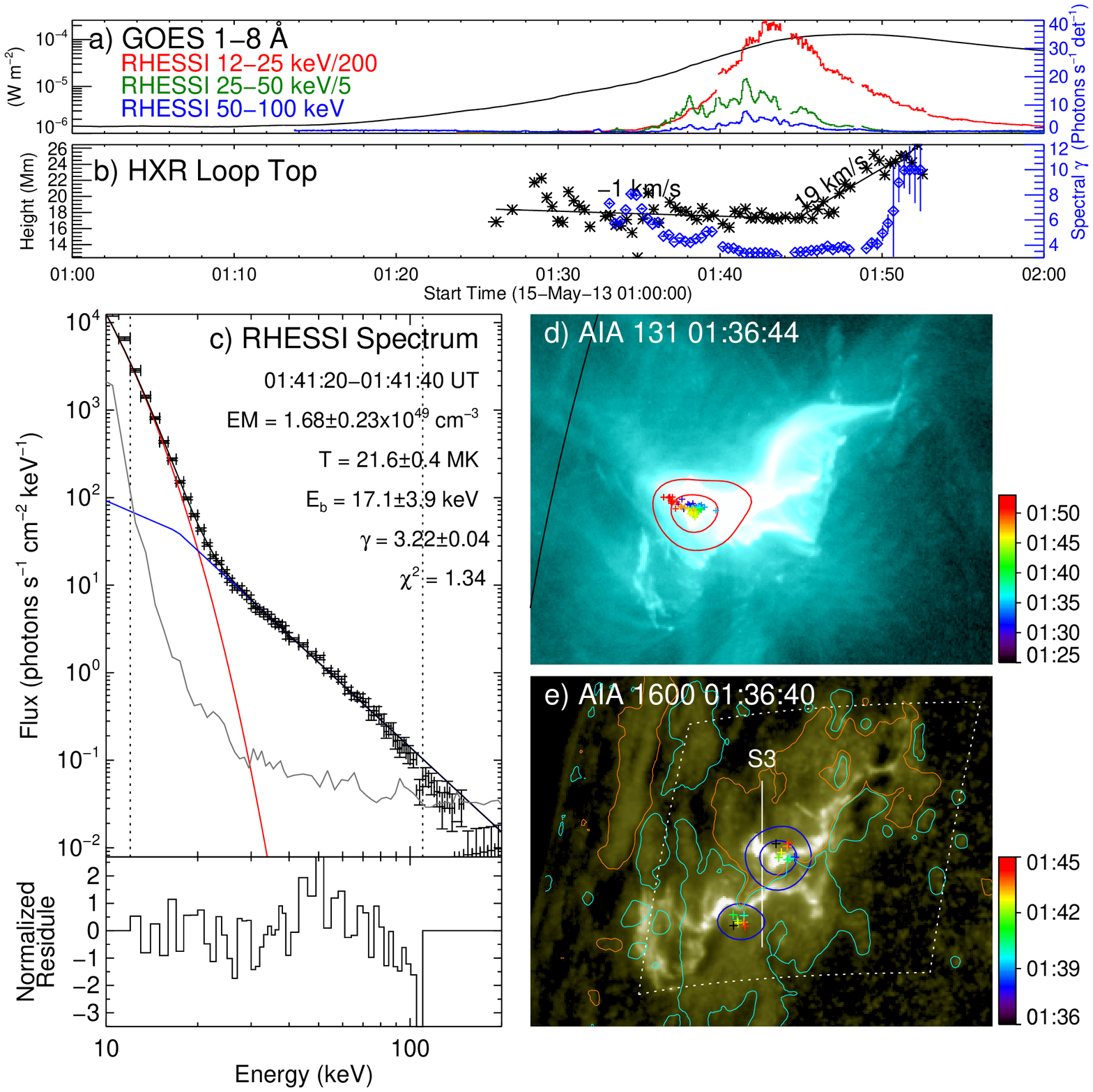}
  \caption{\small Flare spectroscopy and morphology. Panel (a) displays \sat{goes} 1--8~{\AA} flux (black), scaled by the left y-axis, and \sat{rhessi} light curves at 12--25 (red), 25--50 (dark green) and 50--100 keV (blue), scaled by the right y-axis. (b) The projected height of the HXR looptop source (black asterisks) and the HXR spectral index ($\gamma$; blue diamonds). (c) An HXR spectrum at the flare peak is fitted between 12--110 keV (denoted by dotted lines) with an exponential thermal function (red) and a broken power-law function (blue) with $\gamma$ below the broken energy $E_b$ being fixed at 1.5. The background is shown in gray. (d) The flaring loops in AIA 131~{\AA}. (e) The flare ribbons in AIA 1600~{\AA}. Orange (cyan) contours show local $B_z$ at 100 (-100) G. The warped rectangle indicates the FOV of the remapped magnetogram in Fig.~\ref{hmi}. The HXR loop top at 12-15 keV and footpoints at 50-80 keV at the flare peak are demonstrated in red and blue contours (30\% and 70\% of the maximum brightness), respectively. The centroid positions of HXR sources are denoted by plus signs and their temporal evolution is color coded. \label{spectra}}
\end{figure*}

\begin{figure*}
  \centering
  \includegraphics[width=0.85\hsize]{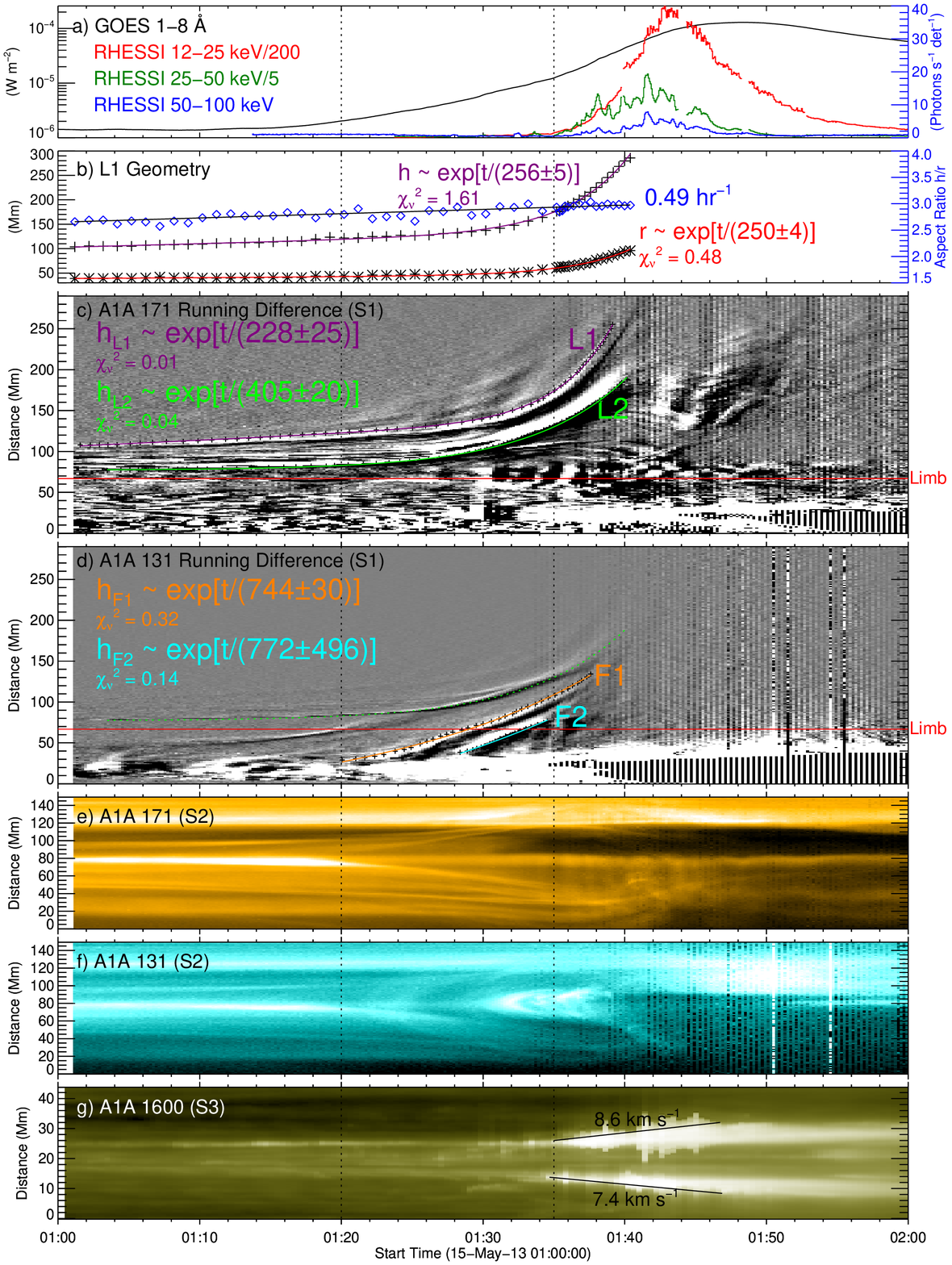}
  \caption{\small Temporal evolution of the eruption. Panel (a) is reproduced from Fig.~\ref{spectra}(a). The onsets of the SXR flare and the nonthermal HXR bursts are indicated by two vertical dotted lines, respectively. Panel (b) shows L1's height ($h$; pluses) and radius ($r$; asterisks) as well as the aspect ratio ($h/r$; diamonds). The dynamic evolution in 171, 131, and 1600~{\AA} as seen through the slits S1, S2 (\fig{overview}(a)) and S3 (\fig{spectra}(e)) are shown in (c)--(g), respectively. The horizontal red line in (c) and (d) marks the limb position of S1. Fittings of the visually picked points on the edges (plus signs) of L1, L2, F1 and F2 are shown in magenta, green, orange and cyan, respectively. The reduced chi-square ($\chi_\nu^2$) as normalized by the degrees of freedom for the fittings are also given. The fitting curve for L2 in (c) is reproduced in (d) in a dotted line.\label{ht}}
\end{figure*}

The flare morphology and spectra as revealed by \sat{rhessi} are typical of a ``standard'' flare. The integrated spectrum over 20 s around the peak of the nonthermal HXR bursts (\fig{spectra}(c)) can be well fitted with an exponential thermal function (red) and a nonthermal power-law function (blue). The thermal component comes mainly from the loop-top source (red contours; \fig{spectra}(d)) and the nonthermal component from a pair of conjugate footpoints (blue contours; \fig{spectra}(e)) located at opposite polarities of the local $B_z$ (orange and cyan contours; see also Fig.~\ref{hmi}). The footpoint sources are associated with two separating UV flare ribbons (\fig{spectra}(e)) and \fig{ht}(g)), with the former being very compact and localized relative to the latter. The nonthermal HXR bursts start at about 01:35 UT, when the corresponding spectral index decreases below $\sim\,$6. Starting from this moment, there is also a clear trend that the height of the flare loop-top source gradually decreases at \aspeed{1} (\fig{spectra}(b) and (d)). Prior to the onset of the nonthermal bursts, the loop-top height is more scattered. This height is calculated as the projected distance between the centroid position (80\% contour level) of the loop-top source and the middle of the centroid positions (50\% contour level) of the flare-peak footpoints (\fig{spectra}(e)). The downward motion of the flare loop-top source persists until the end of the nonthermal bursts, when the loop top source moves upward at a speed of \aspeed{19}, much faster than the downward speed.

To study in detail the evolution of the loop system, two virtual slits are placed intersecting the loop apex (labeled `S1') and across the loop legs (labeled `S2'), respectively, as shown in \fig{overview}(a). A third slit, `S3', is placed across the two flare ribbon in AIA 1600~{\AA} (\fig{spectra}(e)). The lower end of S1 is registered with the flare kernel. In the resultant stack plot generated with S1 in a running-difference approach, loop bundles, L1, L2, and hot channels, F1 and F2, can be readily recognized (\fig{ht}(c) and (d)). In \fig{ht} (e)--(g) stack plots are generated from original images and shown in a logarithmic scale. Through the slit S2, one can see that in AIA 171~{\AA} L1's two legs start to approach each other (\fig{ht}(e)) at the onset of the nonthermal HXR bursts. Meanwhile, the hot channels as detected in 131~{\AA} begin to expand (\fig{ht}(f)). Through the slit S3, one can see that the flare ribbons is explosively enhanced and start to separate at \aspeed{10} at the onset of nonthermal HXRs (\fig{ht}(g)). It is noteworthy that the ribbon in the north starts to brighten even before the onset of the soft X-ray (SXR) flare at 01:20 UT, suggestive of energy release in photosphere/chromosphere.

We use an exponential function plus a linear term, 
\begin{equation}
h=h_0+v_0t+ce^{t/\tau}, \label{eq:exp}
\end{equation} 
to fit the visually picked points (shown in plus signs in \fig{ht}) along the leading edge of the tracks left by L1 and F1 in the stack plots, and the trailing edge of the tracks by L2 and F2. The selection of the leading or trailing edge depends on which one is more clearly defined. Obviously, $\tau$ is the cardinal parameter to decide how fast the individual structure is accelerated. The fitting results yield that
\begin{equation}
\tau_{L1}<\tau_{L2}<\tau_{F1}\simeq\tau_{F2}, \label{eq:tau}
\end{equation} 
i.e., higher structures are generally accelerated faster than lower ones. However, the tracks left by L2 and F1 converge after the onset of the HXR bursts (Fig.~\ref{ht}(d)). Although both structures have become very diffuse when F1 catches up with L2 at about 01:39 UT in AIA 131~{\AA}, one sees little sign of interaction between them (deformation or brightening due to compression) from the animation accompanying Fig.~\ref{overview}. In light of the STB observations (\fig{overview}), we suggest that the observed catch-up from \sat{sdo}'s perspective is a projection effect and that L1 and L2 are separated overlying loops for F1 and F2, respectively.

\begin{figure*}
  \centering
  \includegraphics[width=\hsize]{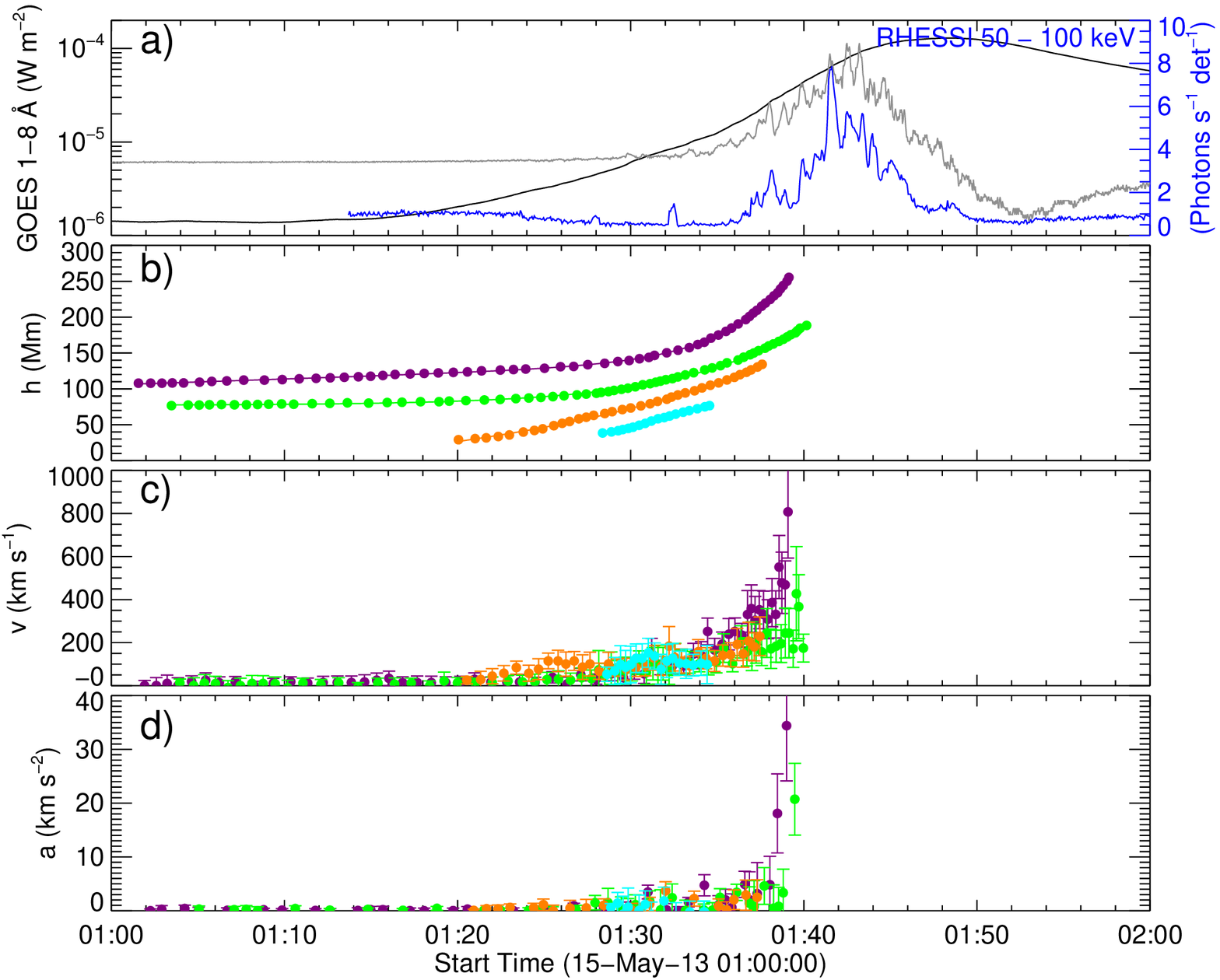}
  \caption{Projected height, velocity and acceleration of L1, L2 and F1, and F2. The same color code as in \fig{ht} is adopted. Panel (a) displays \sat{goes} 1--8~{\AA} flux (black), scaled by the left y-axis, and \sat{rhessi} light curve at 50--100 keV (blue), scaled by the right y-axis. The derivative of the \sat{goes} flux is shown in gray in an arbitrary scale. \label{hva}}
\end{figure*}

The result as implied by Eq.~\ref{eq:tau} is verified independently by deriving the speed and acceleration directly via numeric differentiation (\fig{hva}), using the SolarSoft procedure, \texttt{DERIV\_LUD}, developed by T.~Metcalf. This procedure calculate the derivative by converting an integral representation of the derivative to a matrix equation and solving this by the method of regularization \citep{cb86}. We assume an error of 4 pixels on the measurement of the stack-plot tracks, and the error estimation for the corresponding speed and acceleration is carried out by the procedure in a Monte-Carlo approach, by solving the matrix equation numerous times and recording the rms deviation of the solutions. It is clear from \fig{hva} that L1 rises faster than the other three structures, with a speed up to $808\pm215$ km~s$^{-1}$ and acceleration up to $35\pm10$ km~s$^{-2}$. The CME as observed subsequently by \sat{soho}\footnote{Solar and Heliospheric Observatory} coronagraphs gains an average speed of \speed{1366} (see the SOHO CME catalog\footnote{http://cdaw.gsfc.nasa.gov/CME\_list/}).

The height-time evolution of L1 is studied in an alternative approach, by fitting visually picked points on the rim of L1's upper section with a circle (see \fig{overview}(a) and (b)). The projected height ($h$) is calculated as the distance from the center of the circle to the middle of the footpoint centroid positions at the flare peak (\fig{spectra}(e)) plus the radius. The evolution of $h$ and the radius $r$ are again fit with Eq.~\ref{eq:exp} (\fig{ht}(b)). The fitting of $h$ gives a similar result as fitting L1's leading edge in the stack plot (\fig{ht}(c)). Both $h$ and $r$ follow a similar exponential growth rate, but the aspect ratio $h/r$ increases gradually with time, from 2.66 at 01:01:11 UT to 2.97 at 01:40:23 UT, at an average rate of 0.49 hr$^{-1}$ (\fig{ht}(b)). 

\subsection{Waves \& Oscillations} \label{ss-wave}
\begin{figure*}
  \centering
  \includegraphics[width=\hsize]{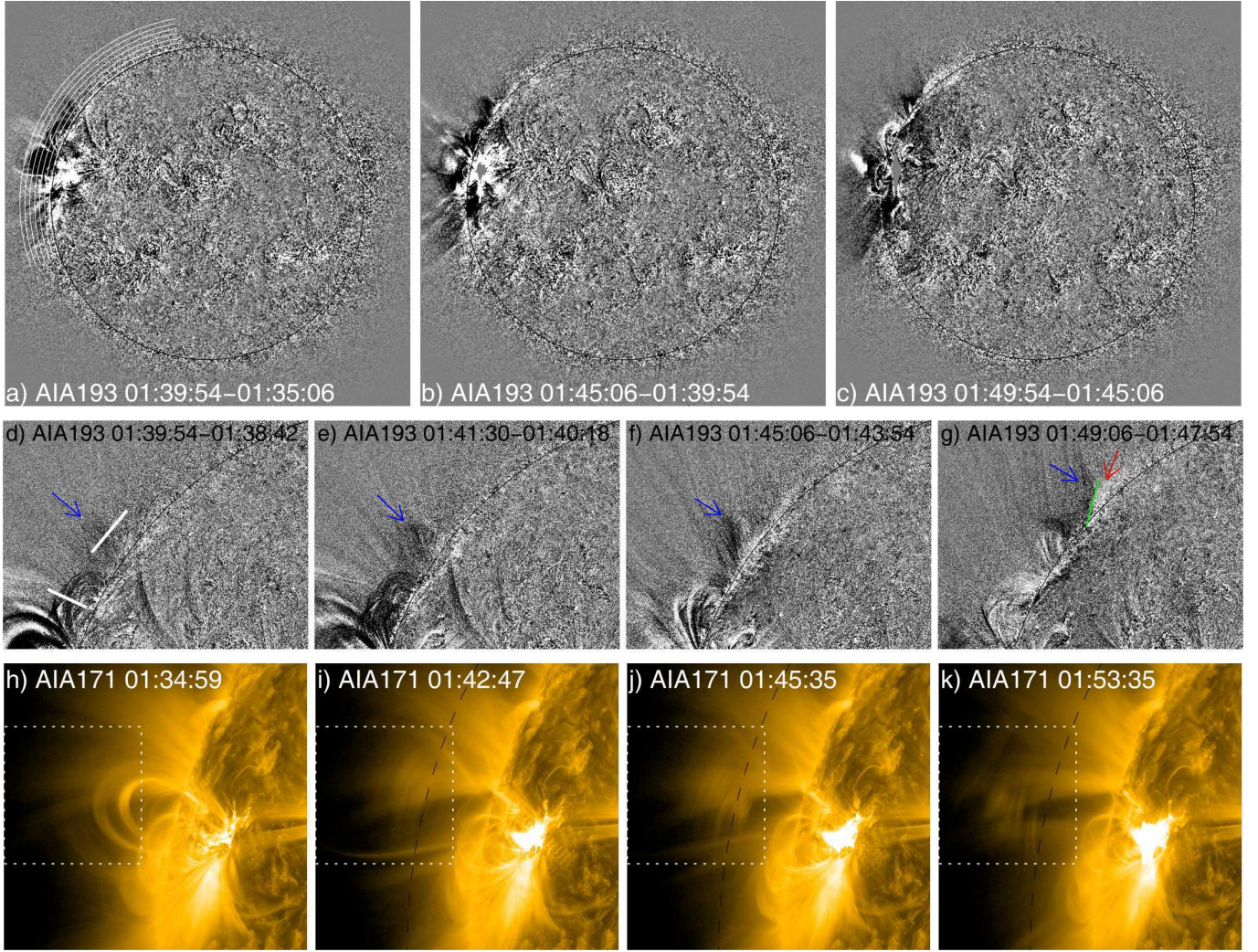}
  \caption{\small Snapshot of waves and oscillations assoicated with the CME. (a--c) AIA 193~{\AA} running difference images showing a global EUV wave propagating to the polar region. A series of arc-shaped slits above the limb are shown in Panel (a). (d--g)  A sequence of AIA 193~{\AA} running difference images showing the rarefaction (darkened) and compression (enhanced) wavefronts as marked by blue and red arrows, respectively. The interface between them is delineated by a green line in Panel (g). The virtual slits adopted to study the oscillations of coronal loops and polar plumes are indicated in (d) by a vertical and horizontal line, respectively. (h--k) A sequence of AIA 171~{\AA} images showing a series of parallel strips underneath the expanding loop system. A reference circle above the limb demonstrates its initial sunward movement and the subsequent outward propagation. The rectangles marks the region used to construct the $k$-$\nu$ diagrams in Fig.~\ref{kw}. An animation of 193 and 171~{\AA} images is available online. \label{wave}}  
\end{figure*}

\begin{figure*}
  \centering
  \includegraphics[width=\hsize]{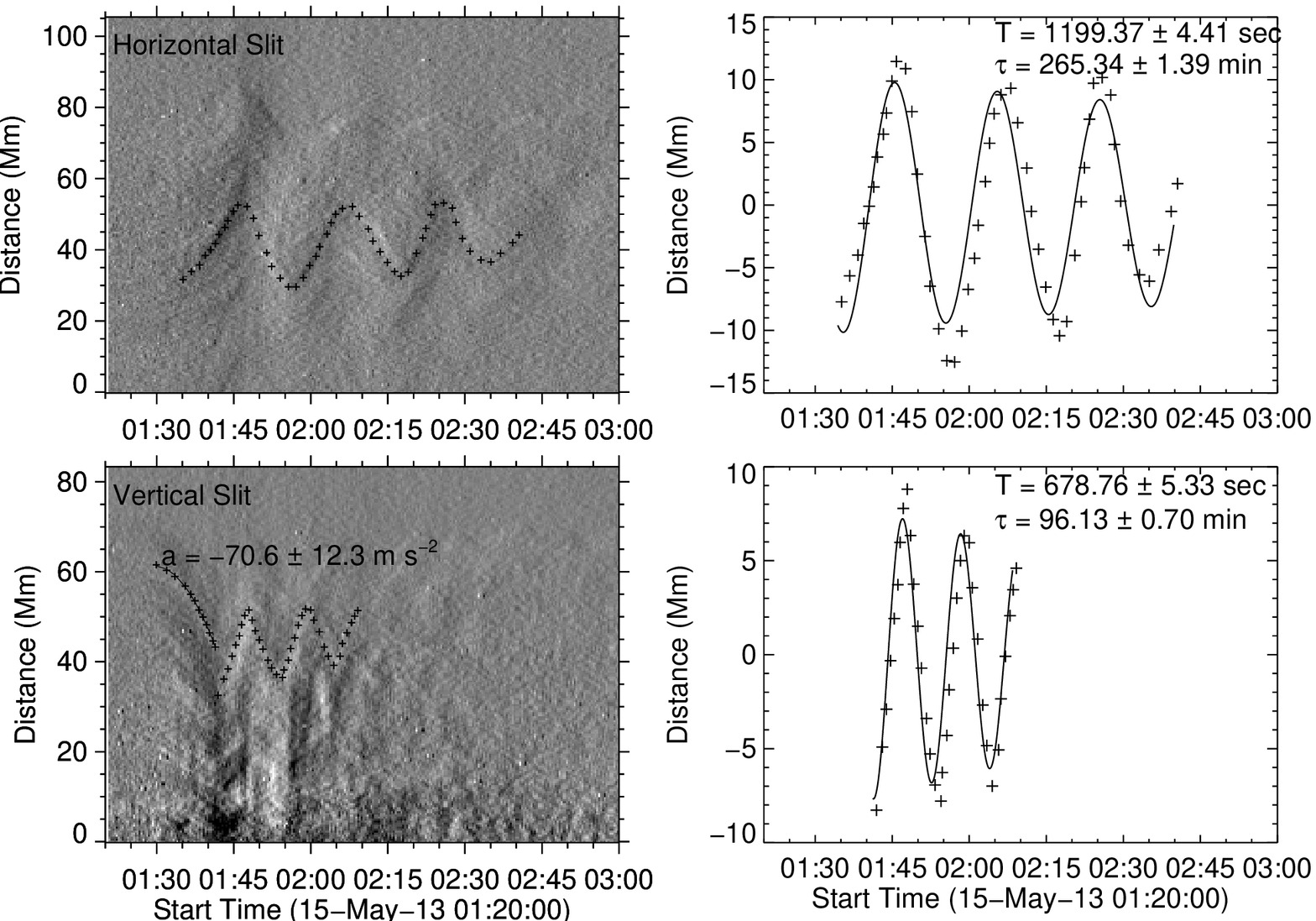}
  \caption{\small Oscillations of polar plumes (top) and coronal loops (bottom) associated with the CME. Left column: stack plots generated through the virtual slits indicated in Fig.~\ref{wave}(d). Visually picked points are indicated by plus signs. Right column: oscillatory patterns identified in the left column are de-trended and fitted with a damped cosine function (Eq.~\ref{eq:cos}).   \label{osci}}  
\end{figure*}

Multiple waves and oscillations are excited by this energetic event, including a global EUV wave (\S\ref{ss-global-wave}) that propagates to as far as the polar region and excites oscillations of polar plumes on its way, and a local EUV wave (\S\ref{ss-local-wave}) underneath the expanding loop system. The former is associated with a metric type II radio burst (\S\ref{ss-type2}). 

\subsubsection{Global EUV Wave} \label{ss-global-wave}
From the AIA 193~{\AA} running-difference images, one can see a wavefront originating from the periphery of the expanding loop system and propagating along the limb toward the north pole (Fig.~\ref{wave}(a--c); see also the accompanying animation). Its southward traveling is less obvious in SDO/AIA observations but quite visible in STB/EUVI images (not shown here). At about 01:30 UT, the expanding loop system (visible at the lower-left corner of Fig.~\ref{wave}(d)) starts to push downward a group of loops to the immediate north of the target active region, and at about 01:40 UT, these loops begin to rebound and subsequently oscillate. A bundle of polar plumes to the north of the oscillating loops also oscillate at approximately the same time, presumably due to the wave propagating through them. A vertical and horizontal virtual slit are placed across the oscillating loops and plumes, respectively (Fig.~\ref{wave}(d)), and the time-distance stack plots obtained from running-difference images are shown in the left column of Fig.\ref{osci}. It is clear that the oscillation of the plumes starts earlier than that of the coronal loops, implying that the wave initiates when the loops are still compressed by the expanding loop system. Both oscillatory patterns are fitted with a damped cosine function (right column of Fig.~\ref{osci}), 
\begin{equation}
f=A\cos\left(\frac{2\pi }{T}t + \phi\right)e^{-t/\tau}, \label{eq:cos}
\end{equation}
and one can see that the two oscillations are distinct from each other in terms of both period $T$ and e-folding damping time $\tau$. 

\begin{figure}
  \centering
  \includegraphics[width=\hsize]{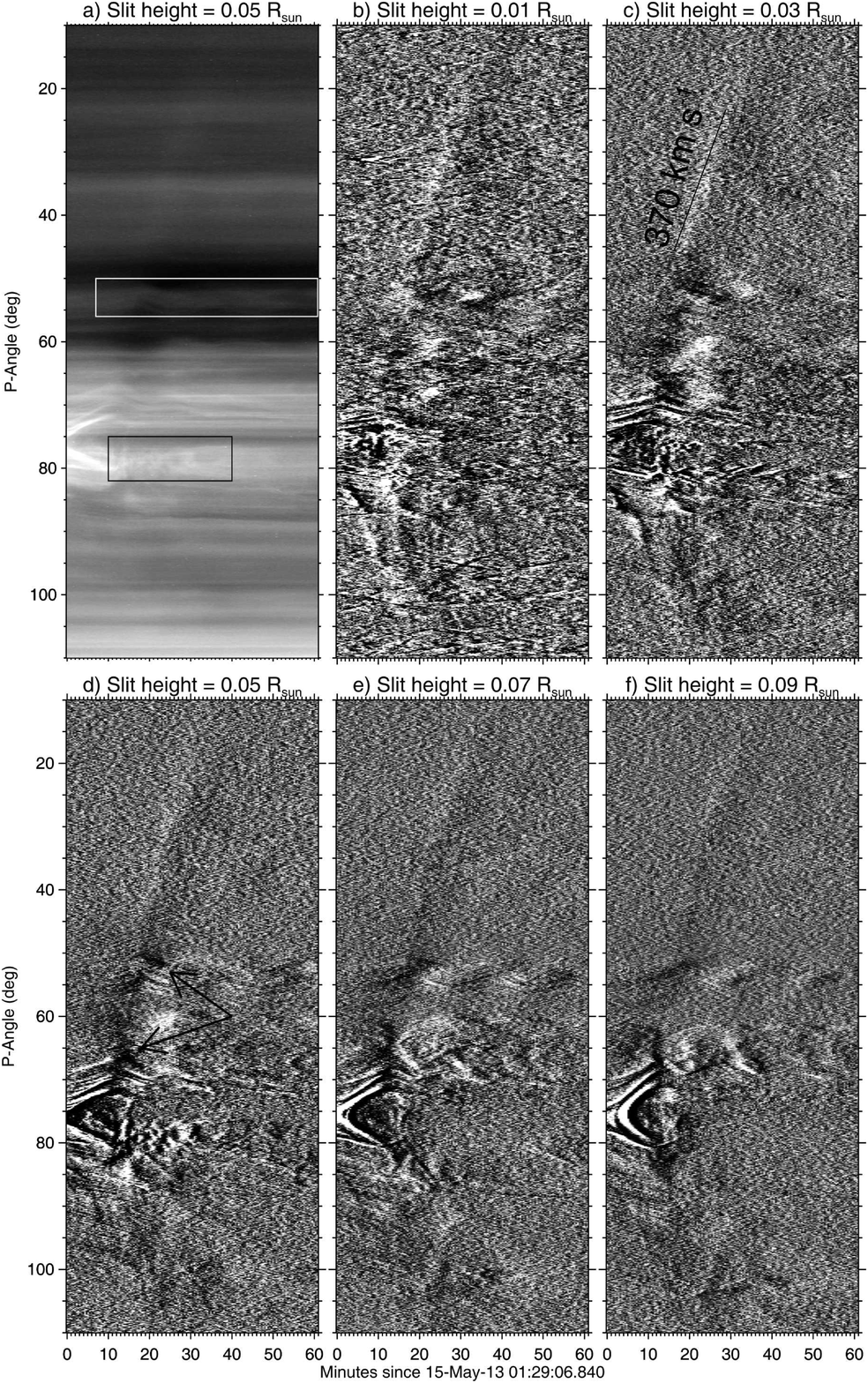}
  \caption{\small Dynamic evolution of low corona seen though the arc-shaped slits above the limb as indicated in  Fig.~\ref{wave}(a). In Panel (a) the stack plot is generated with original AIA 193~{\AA} images, whereas in the rest of panels, the stack plots are constructed with running difference images. The white and black rectangles in (a) mark the p-angle position of the oscillating polar plumes (top panels of Fig.~\ref{osci}) and the local EUV wave train (\S\ref{ss-local-wave}). The blue line in (c) indicates a linear fit to the wavefront track, whose slope yields a speed of \speed{370}. Black arrows in (d) mark the segment of the wavefront track that is rarefaction-dominant (darkened). \label{limbslit}}  
\end{figure}

\begin{figure*}
  \centering
  \includegraphics[width=\hsize]{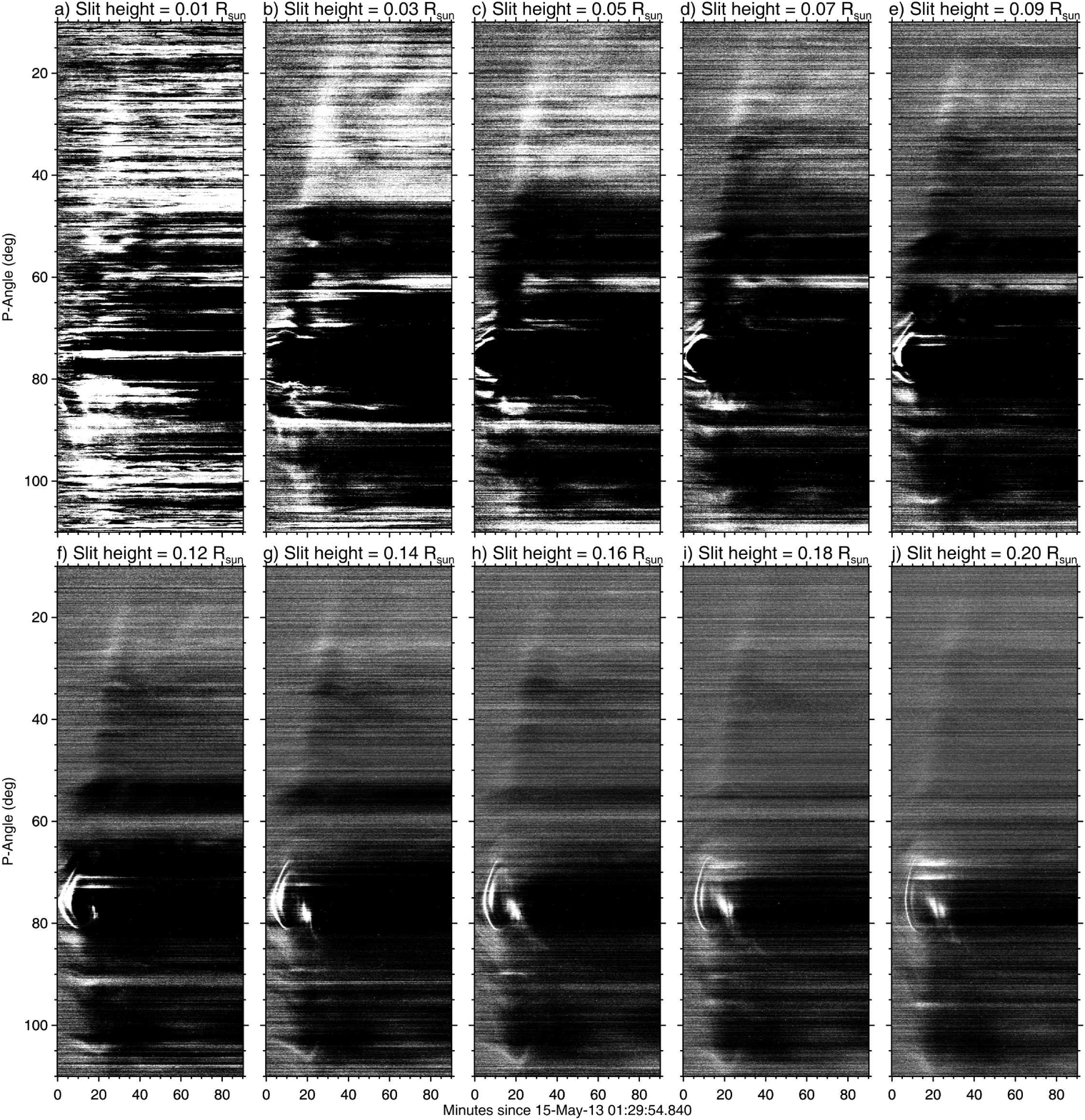}
  \caption{\small Dynamic evolution of low corona seen though a series of arc-shaped slits similar to Fig.~\ref{limbslit}, up to \rsun{0.2} above the limb. The stack plots are generated with base-difference images. \label{base}}  
\end{figure*}

We place a series of arc-shaped slits above the limb to study the propagation of the global EUV wave. These slits span the same p-angle range, from 10 to 110 deg, as displayed in Fig.~\ref{wave}(a). The resultant stack plots obtained from running-difference images are shown in Fig.~\ref{limbslit}. This EUV wave can be detected at various heights from 0.01 up to 0.09 $R_\odot$ above the limb, moving at about \speed{370} in the quiet region. The wavefront from about 01:45 UT onward is characterized by an enhanced track side by side with a darkened one in the stack plots  (Fig.~\ref{limbslit}(b--f)), suggesting that the wave produces compression and then rarefaction when traveling through corona, which is a characteristic of longitudinal waves. Transmitting through polar plumes, the wave must be a fast magnetoacoustic wave, so that it is able to propagate in directions perpendicular to the open fields of the plumes. Earlier in time, the wavefront is dominated by rarefaction, as seen in the bottom panels of Fig.~\ref{limbslit} (marked by arrows in Panel (d)). This is confirmed in running-difference images, in which a darkened wave packet propagating northward is visible as early as 01:40 UT (Fig.~\ref{wave}(d--g); marked by blue arrows). This indicates that initially the wave is not strong enough to produce a visible compression front. At 01:49 UT (Fig.~\ref{wave}(g)) a forward inclined interface (delineated by green line) forms between the darkened and enhanced front (marked by a red arrow). That the coronal wavefront is inclined toward the surface has been reported in the literature as a signature of the increasing Alfv\'{e}n speed with height \citep[e.g.][]{liuw12,liu13}. Base-difference images yield similar stack plots (Fig.~\ref{base}) except that the wavefront is visible up to 0.2 $R_\odot$ above the limb. The rarefaction front can also be seen in these stack plots (e.g., Fig.~\ref{base}(d--g)) as a dimmed extension of the enhanced, compression front back to the source active region, where the coronal dimming developed following the wave becomes dominant.

\begin{figure}
  \centering
  \includegraphics[width=\hsize]{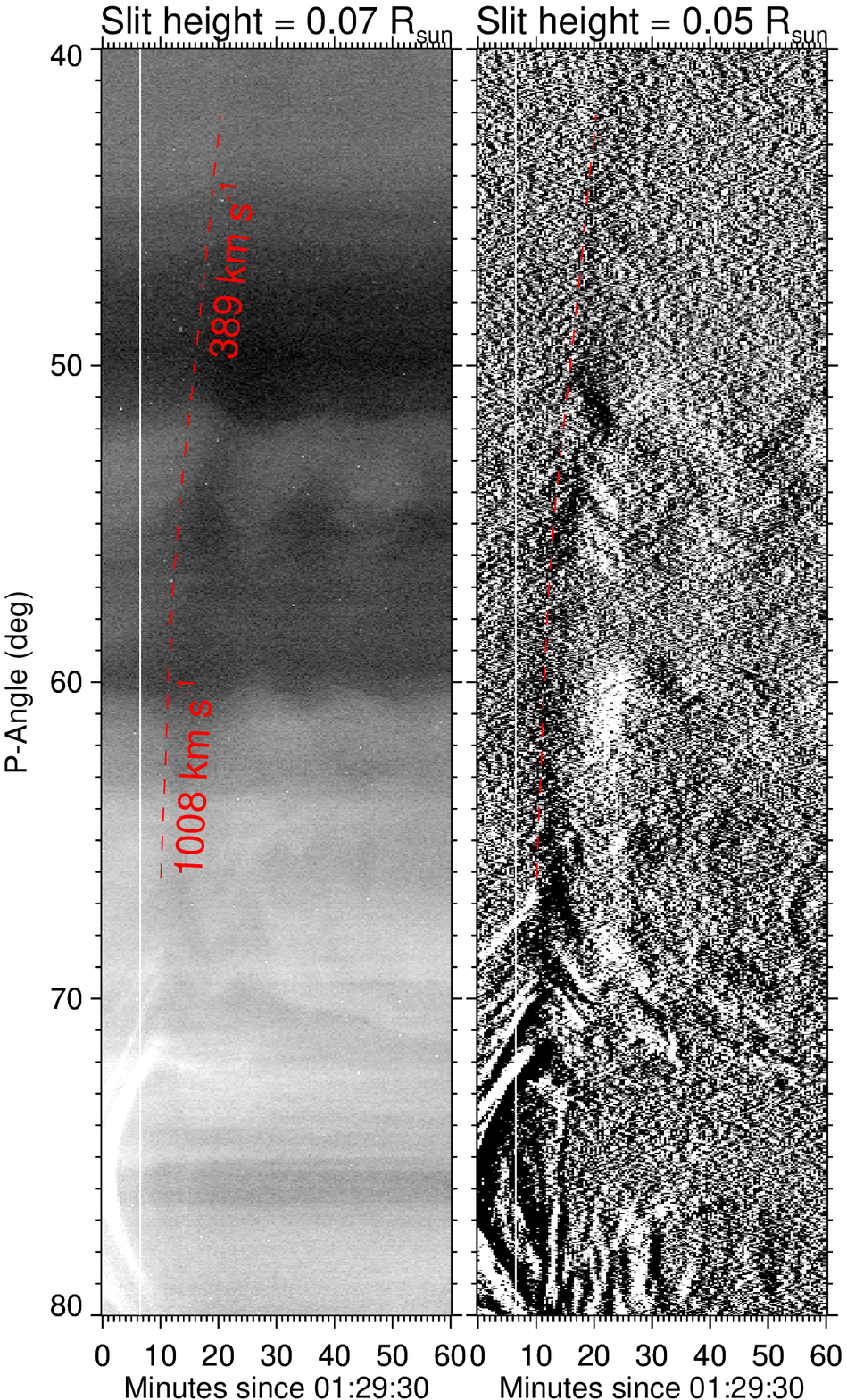}
  \caption{\small Stack plots obtained from two arc-shaped slits at different heights spaning the p-angles from 40 to 80 deg. The ``leading edge'' of the dark, rarefaction front is delineated by the dashed red line in the right panel and replotted in the left panel. Speeds are obtained from piece-wise linear fitting of this curve. The onset of the plume oscillation is indicated by the vertical line. \label{zoom}}  
\end{figure}

We further ``zoom'' into the region with p-angles ranging from 40 to 80 deg to investigate the relationship between the global wave and the oscillating plumes (the latter being marked by a white rectangle in Fig.~\ref{limbslit}(a)). In Fig.~\ref{zoom}, we show a stack plot obtained from the slit at \rsun{0.07} above the limb using original images, which gives the most visible oscillatory pattern of the group of plumes at about 52--55 deg (left panel), and a stack plot from the slit at \rsun{0.05} above the limb using running difference images, which gives the most visible wavefront track (right panel). Despite that the track is quite diffuse, one can still see that the dark, rarefaction front passes through the plumes only slightly earlier than the first peak of the oscillatory pattern, but definitely later than the oscillation onset at about 01:36 UT (marked by a vertical line). This is not unexpected, however, as the oscillating plumes must be responding to the compression front preceding the rarefaction font. 

\subsubsection{Type II Radio Burst} \label{ss-type2}
\begin{figure}
  \centering
  \includegraphics[width=\hsize]{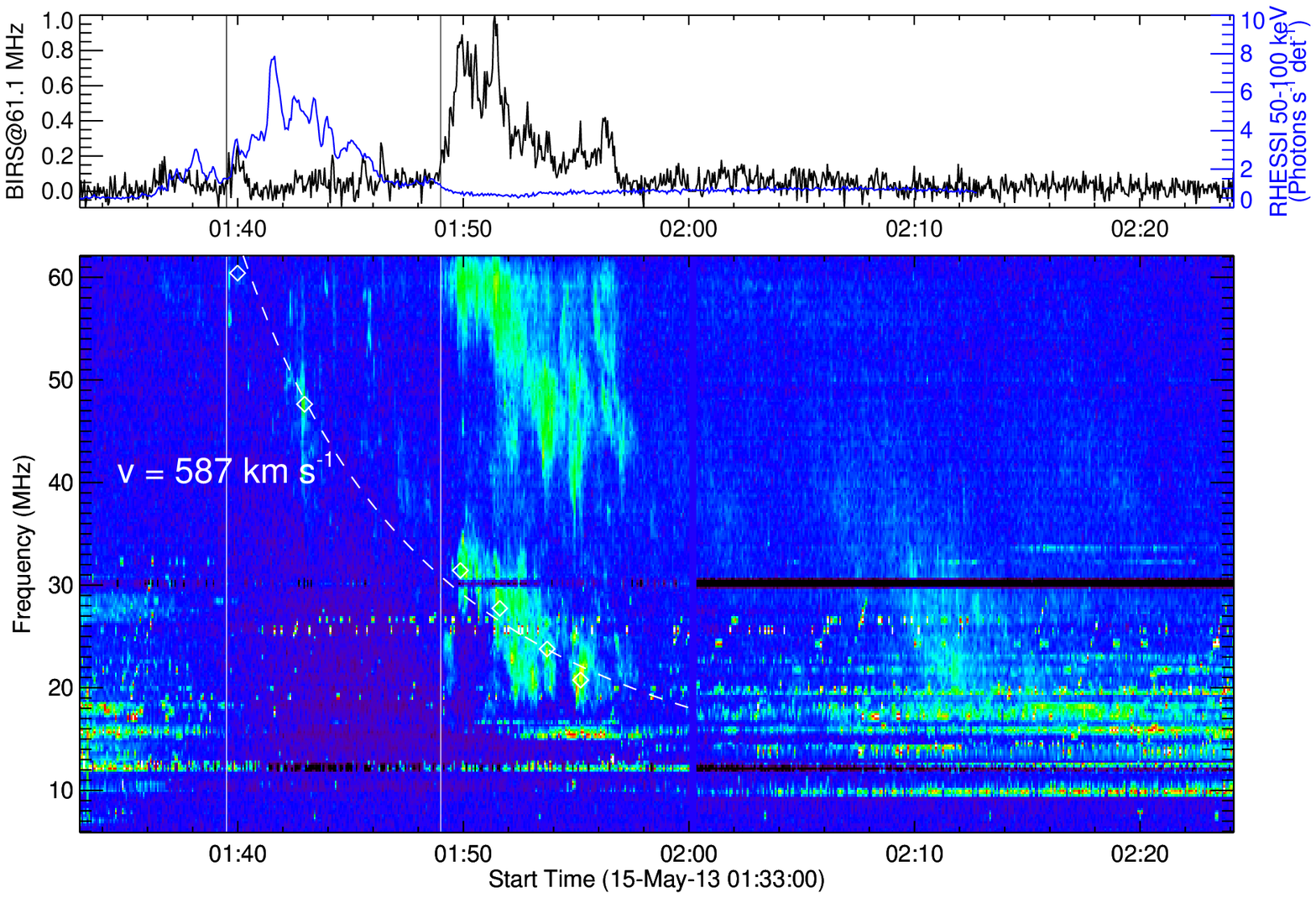}
  \caption{\small A metric type II burst observed by the Bruny Island Radio Spectrometer. Top panel shows the BIRS lightcurve at 61.1 MHz, superimposed with the RHESSI lightcurve at 50--100 keV (blue). Bottom panel displays the dynamic spectrum. The dahsed curve shows the fitting of the visually picked points (diamonds) on the fundamental band using the Newkirk density model. The onset of the type II burst and its later enhancement are marked by two vertical lines. \label{radio}}  
\end{figure}
A metric type II radio burst is observed by the Bruny Island Radio Spectrometer (BIRS) whose frequency range is typically 6 to 62 MHz. The first clear signal of the type II fundamental band appears at 01:39:40 UT, which coincides with the appearance of the rarefaction front in AIA 193~{\AA} running difference images (Fig.~\ref{wave}(d)). Then both the fundamental and second harmonic structures are sharply enhanced from 01:49:00 UT onward, which is coincident with the formation of a clear interface between the rarefaction and compression front (Fig.~\ref{wave}(g)). We adopt the two-fold \citet{newkirk61} density model to fit the hand-picked points along the fundamental band, which yields a propagation speed of \aspeed{580}. It is faster than the speed measured on the compression front after about 01:45 UT through the arc-shaped slits (Fig.~\ref{limbslit}(c)), but considering that large uncertainties are involved in the estimate of the type II speed, this speed difference may not be significant.

\subsubsection{Local EUV Wave} \label{ss-local-wave}
\begin{figure}
  \centering
  \includegraphics[width=\hsize]{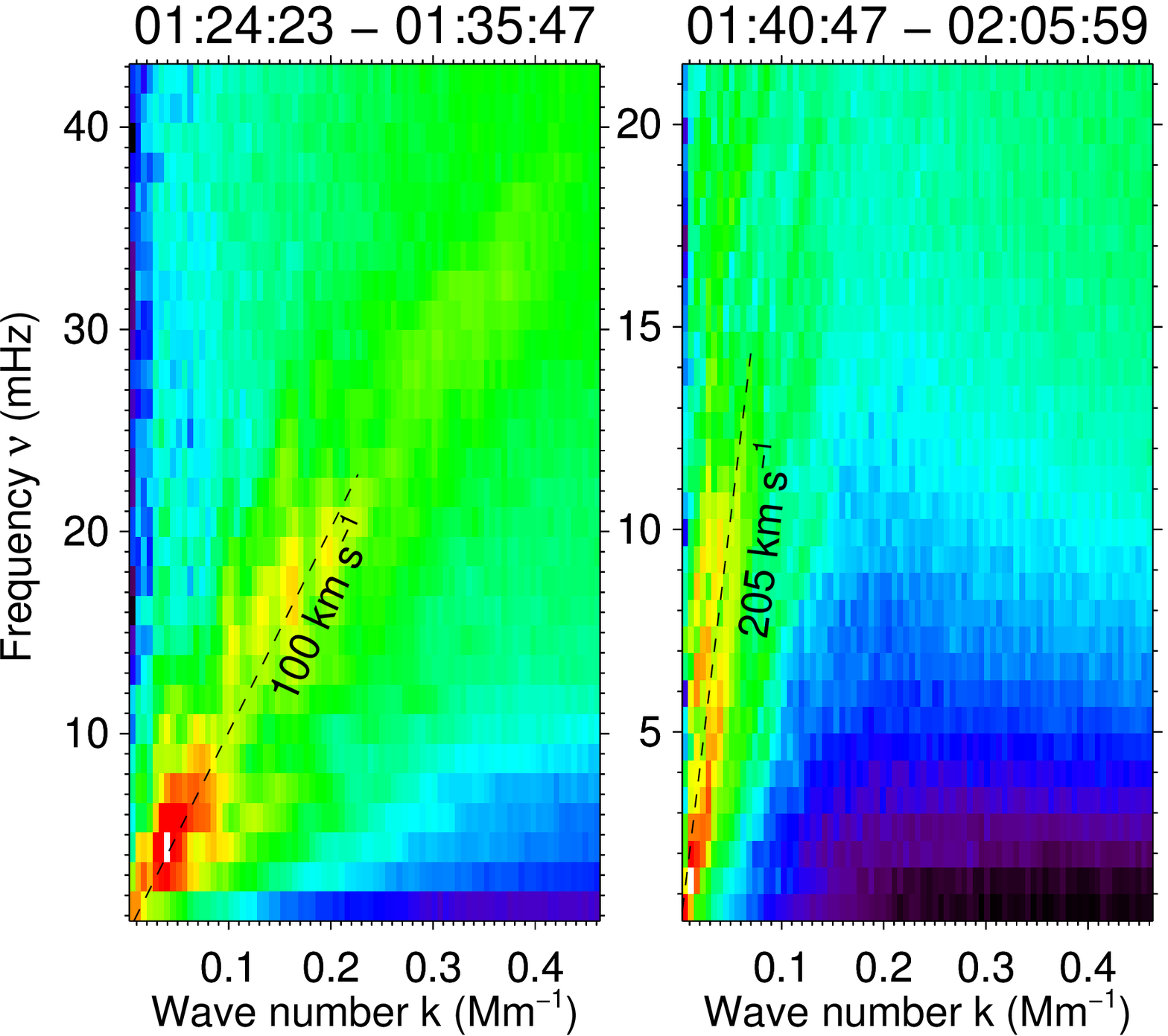}
  \caption{\small $k$-$\nu$ diagrams obtained from the Fourier analysis of the data cube whose spatial dimension is marked by the rectangle in the bottom panels of Fig.~\ref{wave}. The temporal duration of the data cube is indicated by the titles of each panel.\label{kw}}  
\end{figure}
We also notice a series of quasi-periodic strips beneath the expanding loop system, most clearly in AIA 171~{\AA}, but also visible in 131, 193 and 304~{\AA}. These strips, also recorded in the AIA 193~{\AA} stack plot (marked by a black rectangle in Fig.~\ref{limbslit}(a)), have very different appearance than coronal loops. Moreover, they first move sunward during 01:40--01:45 UT and then anti-sunward afterwards until out of the AIA FOV at about 02:06 UT. A reference circle (\rsun{0.13} above the limb) concentric to the solar limb is plotted in Fig.~\ref{wave}(i--k) to demonstrate the reversal of the propagative direction. We extract a data cube from the AIA 171~{\AA} image sequence, whose spatial dimension is marked by rectangles in Fig.~\ref{wave}(h--k), to construct $k$-$\nu$ diagrams via three dimensional Fourier transform followed by summing the Fourier power in the azimuthal direction in the $\{k_x, k_y, \nu\}$ space. During 01:24:23--01:35:47 UT (the time spacing between consecutive images with an exposure time of 2 s is 12 s), the $k$-$\nu$ diagram (left panel of Fig.~\ref{kw}) features the loop expansion at \aspeed{100}, and during 01:45:47--02:05:59 UT (the time spacing between consecutive images with an exposure time of 2 s is 24 s\footnote{Images taken between 01:35:47-01:40:47 UT are not used for the analysis as the time spacings between consecutive images are not uniform with the 171~{\AA} passband working in a flare mode.}), the $k$-$\nu$ diagram (right panel of Fig.~\ref{kw}) demonstrates the propagation of the wave train at \aspeed{200}. Since its phase speed $v_p$ is significantly faster than the typical sound speed in the 0.6 MK (peak response temperature of the 171~{\AA} passband) plasma, $c_s=\sqrt{\gamma k_B T/m_p}\simeq 0.12T[\mathrm{K}]^{1/2}\simeq 90$ km s$^{-1}$, or the phase speed of slow magnetoacoustic waves \citep[e.g.,][]{liuj12}, we conclude that it is also a fast magnetoacoustic wave.  

\section{Discussion \& Conclusion}

The eruptive process of the CME in low corona clearly has three phases:
\begin{enumerate}
\item \textbf{Pre-flare phase} (01:00--01:20 UT): Both SXR and HXR fluxes are at background level. The overlying loop bundles, L1 and L2, are undergoing gradual expansion at several kilometers per second in the wake of a jet-like event occurring in the periphery of the active region. There is no hot channel in corona, but weak surface brightening has been detected in UV, suggestive of photospheric/chromospheric reconnection. 
\item \textbf{Flare thermal phase} (01:20--01:35 UT): SXR fluxes increase, while nonthermal HXRs ($\gtrsim25$ keV) maintain at background level. The overlying loop bundles are undergoing slow expansion at tens of kilometers per second. A double hot channel rises into the corona at similar speeds. The UV ribbon brightening continues to enhance, but no obvious ribbon separation is detected. 
\item \textbf{Flare nonthermal phase} (01:35--01:50 UT): Nonthermal HXR bursts are detected. Both the overlying loop bundles and the hot channels are undergoing impulsive acceleration. The lower section of the legs of the overlying loop bundles approach each other, whereas a pair of UV ribbons move away from each other. The downward motion of the HXR looptop source is apparently associated with the hardening of the HXR spectrum, while the upward motion with the softening of the spectrum, in agreement with \citet{sui06}. An order of magnitude estimation gives the reconnection electric field $E=uB_z>100$ V~m$^{-1}$ \citep[e.g.,][]{lw09}, where the ribbon motion speed $u\sim 10$ km~s$^{-1}$ (see Fig.~\ref{ht}(g)), and $B_z> 100$ G (see Fig.~\ref{hmi}(a)). Electrons accelerated by this strong electric field could be responsible for the observed HXR footpoint emission. 
\end{enumerate}

While the observations are generally consistent with the standard model \citep{kp76,tsuneta97,forbes00}, some unique features are uncovered owing to the high-cadence, high-resolution \sat{sdo} data. These features can be grouped into four aspects and their implications are discussed below.

\subsection{CME Progenitor}
It is the first time that a double hot channel (DHC) is observed. It is unclear whether the DHC consists of two flux ropes, i.e., a double-deck configuration \citep{liu12,cheng14,kliem14}, or a single flux rope with braided branches \citep[e.g.,][]{liu14}. In either case, the two hot channels may be regarded as a single coherent structure, as evidenced by their strikingly similarity in kinematic profiles (Fig.~\ref{ht}).

Before the emergence of the DHC, the reverse S-shaped dimming and its subsequent evolution into a rising hot arch in AIA 94~{\AA} is reminiscent of the transformation from an S-shaped loop into arch-shaped at the onset of the sigmoidal eruption studied by \citet{liu10}. We speculate that the interpretation of the S-shaped loop as a flux rope formed prior to the eruption also applies to the present case. 

\subsection{CME Kinematics}
The CME undergoes an extremely fast acceleration during the nonthermal phase, up to 35 km~s$^{-2}$ in low corona, which is two orders of magnitude larger than the gravitational acceleration on the solar surface, and to our knowledge is the fastest CME acceleration recorded in the literature so far. Previous studies using Yohkoh, SOHO, and STEREO data to measure various features of CMEs all give peak accelerations below 10 km~s$^{-2}$ \citep[e.g.,][]{alex02,gallgher03,zd06,vrsnak07,temmer08,temmer10,bein11,bein12}. It has also been demonstrated that a close temporal correlation exists between the CME velocity and SXRs, as well as between the CME acceleration and the derivative of SXRs, or, nonthermal HXRs, due to the Neupert effect \citep{zhang01,zhang04,temmer08,temmer10}. Like HXR bursts, ``spikes'' in the acceleration-time profile could be smoothed out with low cadence measurements. In our case, the high value of acceleration is measured close to the peak of nonthermal HXRs, hence is expected to decrease soon. The CME acceleration must have ceased by the end of the nonthermal phase at 01:50 UT (Fig.\ref{hva}(a)), judging from the \sat{SOHO} coronagraph measurements from 01:48 UT onward, which is characterized by a deceleration rate of 0.05 km~s$^{-2}$ in a 2nd-order polynomial fit (see again the SOHO CME catalog).

On the other hand, despite that the CME ``under-expands'' laterally in low corona, displaying a slightly increasing aspect ratio $h/r$ with time, both a global and a local EUV wave are detected in the present case (\S\ref{ss-wave}). This is in contrast to the recorded lateral ``overexpansion'' in several studies involving EUV waves \citep[e.g.,][]{pvk10,pvs10}, in which a temporary decrease in the aspect ratio is suggested to be the potential trigger of the waves.

\subsection{Wave Initiation}
It is difficult to determine exactly when the global EUV wave initiates, but one knows for sure that it must be excited earlier than the onset of the plume oscillation at about 01:36 UT. It would take about 6 minutes for the wavefront to travel from the edge of the loop system (p-angle $\sim70$ deg) to the plumes (p-angle $\sim55$ deg) at \aspeed{600}, as suggested by the type II burst. This corresponds well with the time that the expanding loop system starts to push downward the neighboring loops (bottom panels in Fig.~\ref{osci}). We speculate that this downward push produces pressure pulses propagating outward at the local fast magnetoascoutic speed. These pulses later evolve into a wave packet as seen in AIA 193~{\AA} running difference images (Fig.~\ref{wave}(d--h)) and excite weak type II radio burst from about 01:40 UT onward (Fig.~\ref{radio}). At about 01:49 UT, the wave packet might steepen into a shock, as evidenced by the clear interface between the rarefaction and compression front (Fig.~\ref{wave}(g)), which corresponds to the drastic enhancement of the type II burst from that time onward (Fig.~\ref{radio}). 

From a case study, \citet{liuw12} concluded as well that the downward compression may be key in driving the global EUV wave. In our case, the downward compression is vividly manifested as the flattening of coronal loops, whose track on the stack plot can be well fitted by a parabola with a constant acceleration of about \accel{70} (bottom left panel of Fig.~\ref{osci}), suggesting that the total force applied on these loops is more or less steady. However, the pressure exerted by the expanding loop system must increase as the resistance is also expected to increase when the loops are pushed toward the surface. The rebound and subsequent oscillation of these loops indicates that the downward compression alleviates with the rapid rising of the loop system. The rapid rising, which is closely associated with the flare nonthermal phase (Fig.~\ref{ht}), actually leads to the convergence of the two legs of the loop system (Fig.~\ref{overview}), therefore less compression on its neighboring loops. 

The quasi-periodic wave train beneath the expanding loop system is observed to first propagate sunward and then anti-sunward at \aspeed{200}, which is distinct from the quasi-periodic fast propagating (QFP) wave trains recently discovered by AIA \citep{liuw11}. Although both are best seen in 171~{\AA} within CME bubbles, QFPs emanate from flare kernels and travel upward at high speeds of 500--2200 km~s$^{-1}$, with numerous pulses following one another \citep{lo14}. The slower wave train in the present case might be excited by an impulsive compression of the overlying field due to the extremely fast upward acceleration of the loop system, which explains why the wave train appears initially at hight altitudes and propagates downward. The subsequent upward propagation indicates a reflection of the wave due to the rapid increase in the Alfv\'{e}nic speed towards the core of the active region.

\subsection{Eruption Mechanism}
The jet-like event that occurs in the periphery of the target active region (Fig.~\ref{init}) may serve to weaken the field confining the loop system through interchange reconnections between closed and open fields, according to the classical reconnection-driven jet model \citep{heyvaerts77,ss00}, therefore resulting in its slow rise.

The DHC rises later and slower than the leading edge of the expanding loop system. Hence, this is not an eruption driven by the hot channel as in the cases studied by \citet{cheng13}. Rather, it is driven by a preexistent, high-lying flux rope associated with the S-shaped dimming. We suggest that the torus instability is the key eruption mechanism. Since the torus instability sets in when the potential field external to the flux rope declines with height at a sufficiently steep rate \citep{tk07,fg07,aulanier10} and this decline rate normally increases with height above active regions \cite[e.g.,][]{liu14,zhang14}, it is expected to see that the high-lying flux rope becomes unstable earlier and hence move faster than the DHC that emerges from the core of the active region (see Eq.~\ref{eq:tau} and Figs.~\ref{ht} and \ref{hva}). 

To conclude, the superior \sat{sdo} data presented here have equipped us with a better understanding on physical properties of the CME's early evolution, which are potentially useful as inputs/constraints for models. In particular, an extremely fast acceleration experienced by the CME is revealed, which is not possible without high-resolution, high-cadence observations. Such a high value of acceleration must not be unique, but can be easily concealed in previous low-cadence data. We conjecture that the compression of coronal plasmas skirting and overlying the expanding loop system, whose aspect ratio $h/r$ increases with time as a result of the rapid upward acceleration, excites the outward-propagating global EUV wave and the sunward-propagating local EUV wave, respectively.

\acknowledgments The authors are grateful to the \sat{sdo}, \sat{stereo}, and \sat{rhessi} consortia for the free access to the data and the development of the data analysis software.  R.L. acknowledges the Thousand Young Talents Program of China, NSFC 41222031 and NSF AGS-1153226. This work was also supported by NSFC 41131065 and 41121003, 973 key project 2011CB811403, CAS Key Research Program KZZD-EW-01-4, the fundamental research funds for the central universities WK2080000031.

\end{document}